\documentclass[12pt]{extarticle}
\pdfoutput=1 
\usepackage{slashed}
\usepackage{latexsym}
\usepackage{amssymb}
\usepackage{amsmath}
\usepackage{graphicx}
\usepackage{arydshln}
\usepackage{adjustbox}
\usepackage{easybmat}
\usepackage{bbm}
\usepackage{cite}
\usepackage{slashed}
\usepackage{bm}
\usepackage{adjustbox}
\usepackage{booktabs}
\usepackage{multirow} 
\usepackage{rotating}
\usepackage{mathtools}
\usepackage{lscape}
\usepackage{hyperref}
\usepackage{bm}
\usepackage{color}
\usepackage{fancybox,framed}
\usepackage{lscape}
\usepackage{multirow}
\usepackage{fancyhdr}

\usepackage[most]{tcolorbox} 

\definecolor{block-gray}{gray}{0.95}

\newtcolorbox{codeSyntax}{
    enhanced,
    frame hidden,
    colback=block-gray,
    boxrule=0pt,
    borderline west={2pt}{0pt}{gray!80!black}
}

\newcommand{\ii}{\ensuremath{\mathrm{i}}}

\begin{document}
\thispagestyle{empty}
\begin{flushright}
\end{flushright}
\vspace{0.8cm}

\begin{center}
{\Large\sc The Landscape of Composite Higgs Models}
\vspace{0.8cm}

\textbf{
Mikael Chala and Renato Fonseca
}\\
\vspace{1.cm}
{\em {Departamento de F\'isica Te\'orica y del Cosmos,
Universidad de Granada, Campus de Fuentenueva, E--18071 Granada, Spain}}
\vspace{0.5cm}
\end{center}
\begin{abstract} 
We classify all different composite Higgs models (CHMs) characterised by the coset space $\mathcal{G}/\cal{H}$ of compact semi-simple Lie groups $\mathcal{G}$ and $\mathcal{H}$ involving up to 13 Nambu-Goldstone bosons (NGBs), together with mild phenomenological constraints. As a byproduct of this work, we prove several simple yet, to the best of our knowledge, mostly unknown results: (1) Under certain conditions, a given set of massless scalars can be UV completed into a CHM in which they arise as NGBs; (2) The set of all CHMs with a fixed number of NGBs is finite, and in particular there are 642 of them with up to 13 massless scalars (factoring out models that differ by extra $U(1)$'s); (3) Any scalar representation of the Standard Model group can be realised within a CHM; (4) Certain symmetries of the scalar sector allowed from the IR perspective are never realised within CHMs. 
On top of this, we make a simple statistical analysis of the landscape of CHMs, determining the frequency of models with scalar singlets, doublets, triplets and other multiplets of the custodial group as well as their multiplicity. We also count the number of models with a symmetric coset.

\end{abstract}

\newpage


\section{Introduction}
Composite Higgs models (CHM)~\cite{Kaplan:1983fs,Kaplan:1983sm,Dimopoulos:1981xc} postulate that the Higgs degrees of freedom, as well as possibly other scalar partners, are composite (pseudo) Nambu-Goldstone bosons (NGBs) arising from the spontaneous breaking $\mathcal{G}\to\mathcal{H}$ produced during the confinement of some new strongly coupled system, presumably at the TeV scale. Thus, CHMs provide, together with Supersymmetry, one of the most appealing solutions to the so-called hierarchy problem; see Refs.~\cite{Contino:2010rs,Panico:2015jxa} for comprehensive reviews. 

In practice, CHMs have been also studied in connection with other problems that the Standard Model (SM) does not solve, for example dark matter~\cite{Frigerio:2012uc,Marzocca:2014msa,Fonseca:2015gva,Chala:2016ykx,Bruggisser:2016ixa,Wu:2017iji,Balkin:2017aep,Ballesteros:2017xeg,Chala:2018qdf,Balkin:2018tma,Ramos:2019qqa,Davoli:2019tpx,Cacciapaglia:2019ixa,Cai:2020njb}, flavour~\cite{Kaplan:1991dc,Agashe:2009di,Redi:2011zi} or baryogenesis~\cite{Espinosa:2011eu,Chala:2016ykx,Chala:2018opy,DeCurtis:2019rxl}. This explains partially why, as of today, most works on CHMs have focused on particular realisations of this paradigm. To the best of our knowledge, the most extensive description of the different cosets allowed in CHMs is given in Ref.~\cite{Bellazzini:2014yua}, which is however far from complete beyond rank 3. (As the authors manifest, they did not intend to be exhaustive in this respect.) Hence, a comprehensive and systematic study of the landscape of CHMs, describing aspects such as the possible symmetries of the NGB sector or the scalar SM multiplets most present in CHMs, among many others, is still lacking. With the current work we aim to fill this gap.

We classify all CHMs based on the coset of compact Lie groups $\mathcal{G}/\mathcal{H}$ with up to 13 NGBs that preserve custodial symmetry and contain no coloured or fractionally charge scalars. In order to scan the enormous landscape of possibilities, we use \texttt{GroupMath}~\cite{Fonseca:2020vke} plus some extra code to handle the specific demands of this endeavour. We note that since the topological properties of the groups are not important, throughout our work we operate with Lie algebras. 

The article is organised as follows. In section~\ref{sec:chms}, we provide a brief introduction to the mathematical structure of CHMs. We address the problem of whether a given set of massless scalars can be UV completed into a CHM in which they arise as NGBs and how to build the corresponding Lagrangian using only IR information. In section~\ref{sec:landscape}, we discuss the aforementioned classification of CHMs, which we provide fully in the ancillary file {\texttt{Landscape\_CHM.txt}, putting special emphasis on the fact that the landscape is finite, on the possibility of embedding any SM-group scalar content in a CHM, as well as on the absence of certain IR symmetries in CHMs. On the basis of these results,  in section~\ref{sec:results} we discuss the frequency of cosets involving certain types of scalar multiplets of the custodial group $SU(2)\times SU(2)$ (singlets, doublets, triplets, etc.), as well as certain properties (such as being symmetric) across the landscape of CHMs. We conclude in section~\ref{sec:conclusions}.

%
%

\section{Composite Higgs models}
\label{sec:chms}
The CHM idea is inspired by the successful understanding of the dynamics of pions on the basis of chiral symmetry breaking in QCD. To a very good extent, the global symmetry of the two-flavour version of QCD is $G=SU(2)_L\times SU(2)_R$, but the quark condensate $\langle\overline{q}q\rangle$ breaks this spontaneously to $H=SU(2)_{L+R}$. The three pions are identified with the NGBs of the coset $G/H$; all other resonances being much heavier~\cite{Contino:2010rs}. The physics of the system is described by a Lagrangian in which the global symmetry is non-linearly realised~\cite{Coleman:1969sm,Callan:1969sn}.

In very much the same vein, CHMs identify the four real degrees of freedom of the Higgs with four NGBs from the spontaneous symmetry breaking $\mathcal{G}\to\mathcal{H}$ in some new strongly interacting sector. The least one can ask of $\mathcal{H}$ is that it contains the SM electroweak group $SU(2)\times U(1)$, with the Higgs transforming as $\mathbf{2}_{\pm 1/2}$. In practice, constraints from violation of custodial symmetry favour an $\mathcal{H}$ containing the larger $SU(2)\times SU(2)$ subgroup, with the Higgs transforming as $(\mathbf{2},\mathbf{2})$. Other phenomenological considerations restrict further the allowed cosets $\mathcal{G}/\mathcal{H}$; for example, the presence of NGBs with fractional electric charges should be avoided.

All in all, we arrive at the following operational definition:

\begin{codeSyntax}
 A composite Higgs model is to be identified by a group $\mathcal{G}$ and a subgroup $\mathcal{H}\supset SU(2)\times SU(2)$, together with embeddings $\varphi_{1}:\mathcal{H}\rightarrow \mathcal{G}$ and $\varphi_{2}:SU(2)\times SU(2)\rightarrow \mathcal{H}$. For simplicity, rather than tracking the functions $\varphi_{1}$ and $\varphi_{2}$, we consider that two models are different if the branching rules of any representations of (i) $\mathcal{G}$ under $\mathcal{H}$ or (ii) $\mathcal{H}$ under $SU(2)\times SU(2)$ are different. We require that 
  $\text{Adj}(\mathcal{G})\to\text{Adj}(\mathcal{H})+ \bm{r}_\mathcal{H}$ under $\mathcal{H}$, with $\bm{r}_\mathcal{H}\to\bm{r}_{SU(2)\times SU(2)}=(\mathbf{2},\mathbf{2})+\cdots$ under $SU(2)\times SU(2)$ and such that no fractional electric charges are present in $\bm{r}_\mathcal{H}$.
\end{codeSyntax}

The decomposition of any $\mathcal{G}$-representation under $\mathcal{H}$ depends only on the coset space $\mathcal{G}/\mathcal{H}$ and it is unchanged if we were to add a factor group to $\mathcal{G}$ that remains unbroken, namely a spectator group. In other words,
we identify a pair of group-subgroup $\left\{ \mathcal{G^{\prime}},\mathcal{H^{\prime}}\right\} $ with $\left\{ \mathcal{G},\mathcal{H}\right\} $ whenever $\mathcal{G^\prime}=\mathcal{G}\times \mathcal{F}$, $\mathcal{H^\prime}=\mathcal{H}\times \mathcal{F}$ and the $\mathcal{F}$ in $\mathcal{H^\prime}$ is trivially embedded in the  $\mathcal{F}$ of $\mathcal{G^\prime}$ (that is, $\varphi_{1}\left(\mathcal{F}\subset H^{\prime}\right)=\mathcal{F}\subset G^{\prime}$). For example, $SU(4)$ can be embedded in $SU(4)_1\times SU(4)_2$ in two different ways. In one case, the embedded $SU(4)$ coincides with $SU(4)_1$ or $SU(4)_2$ (so one of them is a spectator), while in the other case it is the diagonal combination of $SU(4)_{1,2}$. We keep this last possibility and discard the first one, since it is the same as simply having $SU(4)$ break into nothing

Given an explicit representation of $\mathcal{G}$ and a decomposition of the generators into those spanning $\mathcal{H}$, $T^{\widetilde{a}}$, and those parameterising the coset space, $T^{\hat{a}}$, we can trivially build the leading order NGB Lagrangian upon using the CCWZ formalism~\cite{Coleman:1969sm,Callan:1969sn}. It reads:
\begin{align}\label{eq:lag}
 L = \frac{1}{4} f^2 d_\mu^{\,\hat{a}} \, d^{\mu,\hat{a}}\,, 
\end{align}
where $d_\mu^{\,\hat{a}}$ is the projection of the Maurer-Cartan one-form $\omega_\mu$ onto the coset space. Explicitly:
\begin{align}
 \omega_\mu \equiv -\ii  U^{-1} \partial_\mu U = d_{\mu}^{\,\hat{a}} T^{\hat{a}} + T^{\tilde{a}}\,\text{terms}\,,\quad U=\exp{\left(\frac{\ii \Pi_{\hat{a}}T^{\hat{a}}}{f}\right)}\,.\label{eq:MaurerCartan}
\end{align}
The $\Pi_{\hat{a}}$ parameterise the NGBs, while the constant $f$ can be roughly interpreted as the typical scale of the strong sector. (If $\bm{r}_\mathcal{H}$ is reducible, for example $(1,1)+2\times (\bm{2},\bm{2})$ in $SU(4)/SU(2)^2$, there is in general more than one different scale corresponding to the different singlets in the product of two $d_\mu^{\hat{a}}$ symbols.)

One might ask: Given a real representation $R$ of a group $\mathcal{H}$, is it always possible to find $\mathcal{G}$ such that the coset space $\mathcal{G}/\mathcal{H}$ transforms exactly as $R$, i.e. $\bm{r}_\mathcal{H}=R$? The answer is no, and one of the simplest counter-examples (which therefore does not give rise to a CHM) is the representation $\bm{9}$ of $SU(2)$.\footnote{The smaller real and irreducible representations $\bm{r}_\mathcal{H}=\bm{3}$, $\bm{5}$ and $\bm{7}$ can be obtained from the groups $\mathcal{G}=SU(2)\times SU(2)$, $SU(3)$ and $SO(5)$, respectively. } The only group $\mathcal{G}$ such that $\mathcal{G}/SU(2)$ is 9-dimensional is $\mathcal{G}=SU(2)^4$,\footnote{We are disregarding groups with $U(1)$'s since for those the coset space contains at least one singlet of $SU(2)$.} and the associated coset space contains only $SU(2)$ triplets and/or singlets.

However, as we argue later on, for any choice of representation $R_{SU(2) \times SU(2)}$ of the custodial group  it is always possible to pick a $\mathcal{G}$ and a subgroup $\mathcal{H}$ such that $\mathcal{G}/\mathcal{H}$ transforms exactly as $R_{SU(2) \times SU(2)}$.

Equally interesting is the fact that if $R$ fullfills a so-called closure condition, there is at least one group $\mathcal{G}$ such that $\bm{r}_\mathcal{H}=R$. This is best seen by first taking a top-down approach, so let us start by assuming that there is a group $\mathcal{G}$ with generators $T^a$; as above, tildes distinguish those $T$'s which span a subgroup $\mathcal{H}$ and hats stand for the `broken' generators associated to $\mathcal{G}/\mathcal{H}$. The fact that $\mathcal{G}$ is a group means that the commutator of two $T$'s can be expressed as a linear combination of the generators:
\begin{equation}
\left[T^{a},T^{b}\right]=\ii f_{abc}T^{c}\label{eq:commutator}
\end{equation}
for some anti-symmetric structure constants $f_{abc}$. Furthermore, it is well known that this tensor obeys the Jacobi identity
\begin{equation}
f_{iab}f_{icd}+f_{iad}f_{ibc}+f_{iac}f_{ida}=0\,.\label{eq:Jacobi}
\end{equation}
It contains four free indices, and we now consider what happens when each of them is associated to either a preserved or a broken generator of $\mathcal{G}$. Given that permutations of $\left\{ a,b,c,d\right\}$ leave the equation invariant, there are just five possibilities: there can be none, one, two, three or four indices associated with $\mathcal{H}$ (with a tilde), with the rest being associated to $\mathcal{G}/\mathcal{H}$ (with hats). Assuming that the coset $\mathcal{G}/\mathcal{H}$ is symmetric, that is
 \begin{equation}
f_{\hat{a}\hat{b}\hat{c}}=0 \,,
\end{equation}
we get three non-trivial relations from Eq.~\eqref{eq:Jacobi}. 
\begin{enumerate}
	\item The first, involving only $f_{\widetilde{a}\widetilde{b}\widetilde{c}}$, is simply the Jacobi identity for the subgroup $\mathcal{H}$.
	\item Another relation involves $f_{\widetilde{a}\widetilde{b}\widetilde{c}}$ and $f_{\widetilde{a}\hat{b}\hat{c}}$. Note that this last tensor corresponds to the entries of the $\mathcal{H}$-matrices under which the coset space transforms (let us call them $T^{\widetilde{a}}_{\bm{r}_\mathcal{H}}$, so that $f_{\widetilde{a}\hat{b}\hat{c}}=\ii \left(T^{\widetilde{a}}_{\bm{r}_\mathcal{H}}\right)_{\hat{b}\hat{c}}$). With this in mind, the second relation is just Eq.~\eqref{eq:commutator} for the subgroup $\mathcal{H}$ and the $\bm{r}_\mathcal{H}$ representation: $\left[T_{\bm{r}_\mathcal{H}}^{\widetilde{a}},T_{\bm{r}_\mathcal{H}}^{\widetilde{b}}\right]=\ii f_{\widetilde{a}\widetilde{b}\widetilde{c}}T_{\bm{r}_\mathcal{H}}^{\widetilde{c}}$. 
	\item Finally, there is a relation involving only $f_{\widetilde{a}\hat{b}\hat{c}}$, or equivalently $T^{\widetilde{a}}_{\bm{r}_\mathcal{H}}$, which reads $\left(T_{R}^{\widetilde{i}}\right)_{\hat{a}\hat{b}}\left(T_{R}^{\widetilde{i}}\right)_{\hat{c}\hat{d}}+\left(T_{R}^{\widetilde{i}}\right)_{\hat{a}\hat{c}}\left(T_{R}^{\widetilde{i}}\right)_{\hat{d}\hat{b}}+\left(T_{R}^{\widetilde{i}}\right)_{\hat{a}\hat{d}}\left(T_{R}^{\widetilde{i}}\right)_{\hat{b}\hat{c}}=0$. This is the so-called closure condition~\cite{Low:2014nga}.
\end{enumerate} 

Now we can revert this top-down point-of-view and make the following bottom-up observation. By picking any real represent representation $R$, of any group $\mathcal{H}$, that obeys the closure condition, we can extend the structure constants of $\mathcal{H}$. These extended structure constants respect the Jacobi identity, and therefore they are associated to some bigger group $\mathcal{G}$ containing $\mathcal{H}$ and such that the coset $\mathcal{G}/\mathcal{H}$ is symmetric.\footnote{Note that if we find any tensor which respects Eq.~\eqref{eq:Jacobi}, we obtain a new group/algebra. That is because this equation is simply Eq.~\eqref{eq:commutator} applied to $T^a_{bc}=-\ii f_{abc}$.} We wish to emphasize that for this constructive argument one only needs to have the representation matrices of $R$, and in fact  the identity of the group $\mathcal{G}$ built in this way is far from obvious.

For symmetric cosets, the NGB interaction Lagrangian is then completely fixed by the choice of $\mathcal{H}$ and $r_\mathcal{H}$. One can also see that it is so from the CCWZ formalism. Using the notation $\textrm{ad}_{A}\left(B\right)\equiv\left[A,B\right]$, the Maurer-Cartan one-form $\omega_\mu$ in Eq.~\eqref{eq:MaurerCartan} can be written as
\begin{align}
\omega_{\mu} \equiv\omega_{\mu}^a T^{a} & =-\ii\left(\frac{I-\exp\left(-\frac{\ii}{f}\textrm{ad}_{\Pi}\right)}{\textrm{ad}_{\Pi}}\right)\left(\partial_{\mu}\Pi\right)\nonumber \\
& =\sum_{n=0}^{\infty}\frac{\left(-\ii\right)^{n}}{f^{n+1}\left(n+1\right)!}\left(\textrm{ad}_{\Pi}\right)^{n}\left(\partial_{\mu}\Pi\right)\nonumber \\
& =\frac{1}{f}\partial_{\mu}\Pi-\frac{\ii}{2f^{2}}\left[\Pi,\partial_{\mu}\Pi\right]-\frac{1}{6f^{3}}\left[\Pi,\left[\Pi,\partial_{\mu}\Pi\right]\right]+\cdots\,
\end{align}
with $\Pi \equiv \Pi_{\hat{a}}T^{\hat{a}}$. The infinite tower of commutators in this expression is removable with the help of Eq.~\eqref{eq:commutator}:
\begin{equation}
\omega_{\mu}^a=-\ii\left[\mathcal{F}^{-1}\left(I-\exp\left(-\frac{\ii}{f}\mathcal{F}\right)\right)\right]_{a\hat{b}}\left(\partial_{\mu}\Pi\right)_{\hat{b}}\,,
\end{equation}
where $\mathcal{F}_{ab}\equiv \ii\Pi_{\widehat{x}}f_{\widehat{x}ba}$ is just a matrix. When the index $a$ is associated to one of the broken generators, $\omega_{\mu}^{a}$ is nothing but the $d_{\mu}^{\hat{a}}$ in  Eqs.~\eqref{eq:lag} and \eqref{eq:MaurerCartan}. Furthermore, for a symmetric coset only the $f_{\widetilde{a}\hat{b}\hat{c}}$ structure constants are used inside the matrix $\mathcal{F}$, which becomes off-diagonal. We get that
\begin{equation}
d_{\mu}^{\hat{a}}=\left[\mathcal{F}^{-1}\sin\left(\frac{\mathcal{F}}{f}\right)\right]_{\hat{a}\hat{b}}\left(\partial_{\mu}\Pi\right)_{\hat{b}}\,.\label{eq:8}
\end{equation}
A Taylor expansion of the sine function shows that this expression has no odd powers of $\mathcal{F}$; it is only a function of $\mathcal{F}^2$ which is equivalent to the matrix  $\mathcal{T}$ (times $f^2$) in \cite{Low:2018acv}. That the Lagrangian of a CHM can be built only from IR information when the closure condition holds was first pointed out in Refs.~\cite{Low:2014nga,Low:2014oga} (and investigated further in Refs.~\cite{Low:2018acv,Liu:2018vel}). The exposition in there starts from the observation that the Adler's zero condition~\cite{Adler:1964um}, namely that fact that the amplitudes involving NGBs must vanish when one external momentum is zero, can be taken as the defining property of a NGB, which is thereafter enforced at the level of the Lagrangian. We think that our approach provides a different perspective on this topic. 

\section{The Landscape of composite Higgs models}
\label{sec:landscape}
Our main goal is to find all cosets of up to certain size that are compatible with the definition of a CHM given in the previous section. Before going into specific details, we make the following simple observation:

\begin{codeSyntax}
The set of different composite Higgs models with $m$ Nambu-Goldstone bosons is finite.
\end{codeSyntax}
This holds for the following reason. Given that the number of subgroups $\mathcal{H}$ of a group $\mathcal{G}$ is finite, for a fixed $\mathcal{G}$ the number of cosets $\ensuremath{\mathcal{G}}/\ensuremath{\mathcal{H}}$ with $m$ generators is obviously finite as well. However, one might imagine that for larger and larger $\mathcal{G}$'s one  finds an infinite set of pairs group-subgroup with exactly $m$ broken generators. It so happens that this is not the case: As the size of $\mathcal{G}$ grows, looking at its maximal subgroups we observe that the minimum number of broken generators also increases. For simple groups, this fact is visually depicted on the left plot of Fig.~\ref{fig:groups}; for the infinite classical families $SU(n)$, $Sp(2n)$ and $SO(n)$ the trend exhibited in this figure is true for arbitrary large values of $n$. More generally, when the group is semi-simple, $\mathcal{G}=\mathcal{G}_1 \times \mathcal{G}_2 \times \cdots $, its maximal subgroups are obtained either by taking the maximal subgroup of one of the factors $\mathcal{G}_i$, or by taking the diagonal subgroup of two equal factors, i.e. $\mathcal{G}_{i}\times\mathcal{G}_{i}\rightarrow\mathcal{G}_{i}$. The first possibility has just been addressed, while the second one necessary involves a larger amount of NGBs as the size or number of the $\mathcal{G}_i$ increases (we stress again that in this work we require that none of the $\mathcal{G}_{i}$ factors of $\mathcal{G}$ remains unbroken).

With this said, in this work we consider all CHMs with up to $m=13$ NGBs. Computational time is an important consideration, but it is certainly possible to go beyond this bound; we adopt this limit because the number of CHMs is already quite large for $m\leq13$, allowing us to highlight the importance and feasibility of thoroughly scanning the landscape of possible models.

There are only 18 simple groups $\mathcal{G}$ for which one can have 13 or less NGBs, the largest of which is $SO(14)$; see Fig.~\ref{fig:groups}. This last group has rank 7, however if non-simple groups are considered then we must scan up to rank 8, since the coset $SU(2)^8/SU(2)^4$ is associated to 12 NGBs. In total, ignoring $U(1)$ factors, there are $472$ groups with rank no larger than 8 (see the left plot on Fig.~\ref{fig:groups}).

\begin{figure}[t]
 \includegraphics[width=1.0\columnwidth]{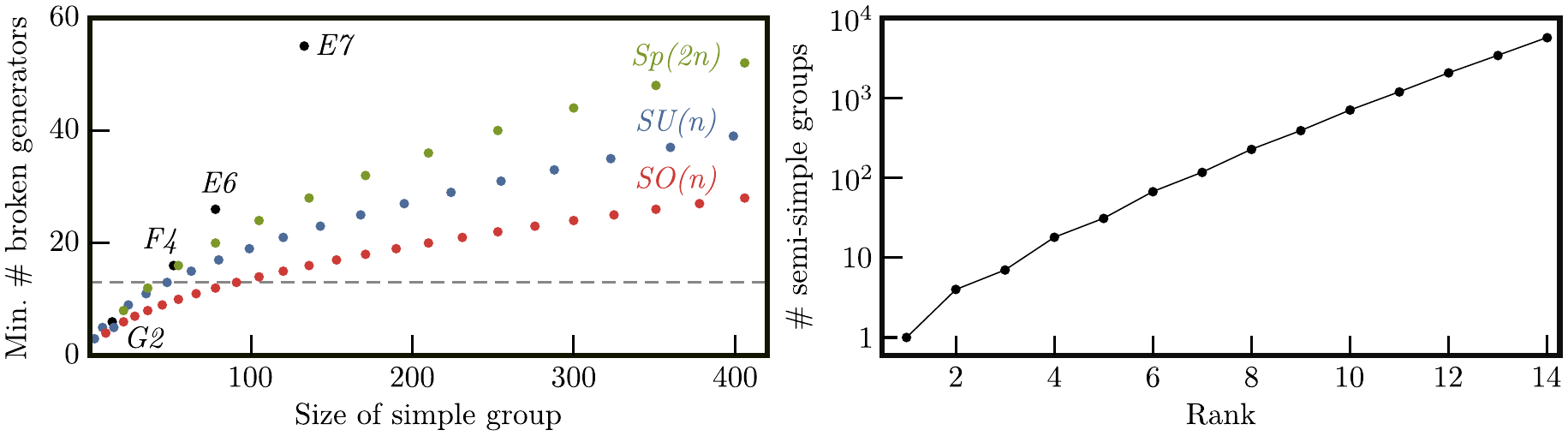}
 \caption{\it Left) Size of simple groups $\mathcal{G}$, and the minimum number of broken generators which can be obtained by looking at the largest subgroup of $\mathcal{G}$ with no abelian factors. 
 	 $E_8$ is not shown (the size of this algebra is 248 and at least 112 generators must be broken). Groups above the dashed line are associated to more than 13 NGBs and therefore are not considered in our scan. (Note that in our scan we also consider semi-simple groups.) Right) Number of semi-simple groups (excluding abelian $U(1)$ factors) of a given rank.}\label{fig:groups}
\end{figure}

For each of these groups, we scan over all possible subgroups and check whether the corresponding coset fullfils the requirements of a CHM, for which we rely on \texttt{GroupMath}~\cite{Fonseca:2020vke}. CHMs differing by extra $U(1)$'s in either $\mathcal{G}$ or $\mathcal{H}$ are trivial to account for. This is because the effect of each extra $U(1)$ in $\mathcal{G}$ is simply to add a NGB transforming as a singlet of $\mathcal{H}$, and as $(1,1)$ under the custodial group $SU(2)\times SU(2)$. As for $\mathcal{H}$, assuming that $\mathcal{H}\times U(1)$ is also a subgroup of $\mathcal{G}$, the coset space $\mathcal{G}/(\mathcal{H}\times U(1))$ has one less singlet NGB in comparison to $\mathcal{G}/\mathcal{H}$. Therefore, the only non-trivial information that needs to be tracked is, for each $\mathcal{G}$ and $\mathcal{H}$ with no abelian factors, what is the maximum number $n_\text{max}$ of $U(1)$'s that can also be left unbroken; $\mathcal{G}/(\mathcal{H}\times U (1)^{n_\text{max}})$ has the lowest number of $\mathcal{H}$ singlets and it is always possible to have more of them upon removing $U(1)$'s from the subgroup or by adding them to $\mathcal{G}$.

A very important point in this systematic exploration of the CHM landscape is the different ways in which the same $\mathcal{H}$ can be embedded into a larger given $\mathcal{G}$. Following Ref.~\cite{Dynkin:1957um}, we consider that two embeddings of $\cal{H}$ into $\cal{G}$ are (linearly) equivalent if they lead to the same branching rules for all representations of $\cal{G}$. In practice though, there is no need to inspect the branching rules of all irreducible representations of $\cal{G}$; only one irreducible representation, or two in the case of $SO(2n)$ groups, must be considered~\cite{Dynkin:1957um,Minchenko:2006}. Furthermore, it is standard practice to disregard embeddings which are obtained trivially from others by some symmetry of $\mathcal{G}$ belonging to the outer automorphism group of $\mathcal{G}$. We  therefore also adopt this practice here. For example, the two embeddings of SU(3) in SU(4) associated to the branching rules $\boldsymbol{4}\rightarrow 1+\boldsymbol{3}$ and $\boldsymbol{4}\rightarrow 1+\overline{\boldsymbol{3}}$ are taken to be the same, since they are converted into one-another via the $Z_2$ automorphism group of $SU(4)$ (which has the effect of conjugating representations).\footnote{A rather technical but nevertheless  important consequence of factoring also these variations is that even for the $SO(2n)$ groups, in order to distinguish two embeddings it is enough to consider how the fundamental representation decomposes under them. }

It is often overlooked that there can be more than one inequivalent embeddings of $\mathcal{H}$ in $\mathcal{G}$, even though their number can be rather large, particularly when the two groups have very different dimensions. For example, it is not hard to prove (see section 3.4 of Ref.~\cite{Fonseca:2015aoa}) that there are $p(n)-1$ inequivalent embeddings of $SU(2)$ in $SU(n)$, where $p(n)$ is the function that counts the distinct number of ways of partitioning the integer $n$ as a sum of positive integers: $p(1), p(2), \cdots = 1,2,3,5,7, \cdots$; $p(n)\sim\exp\left(\pi\sqrt{2n/3}\right)/\left(4\sqrt{3}n\right)$ for large $n$. As a further example, $SU(3)$ and $SU(3)\times SU(2)$ can be embedded in $\mathcal{G}=SU(10)$ in 10 and 35 inequivalent ways, respectively.

The list of all CHMs with up to 13 NGBs can be found in the ancillary material \texttt{Landscape\_CHM.txt}. We find 642 CHMs, provided we do not distinguish those models differing only on $U(1)$ factors.  Each element of the list corresponds to a different CHM, for which the following information is provided: (1) name of $\mathcal{G}$ as a string, (2) name of $\mathcal{H}$ as a string, (3) Cartan matrix of $\mathcal{G}$, (4) Cartan matrix of $\mathcal{H}$, (5) projection matrix of the embedding of $\mathcal{H}$ in $\mathcal{G}$, (6) $\bm{r}_\mathcal{H}$, (7) projection matrix of the embedding of $SU(2)\times SU(2)$ in $\mathcal{H}$, (8) $\bm{r}_{SU(2)\times SU(2)}$. Thus, for the next-to-minimal CHM~\cite{Gripaios:2009pe}, we have:
\begin{codeSyntax}
 %
 \texttt{{\{"SU4"\}},\,\,\,{\{"SO5"\}},\,\,\,{\{\{\{2,-1,0\},\{-1,2,-1\},\{0,-1,2\}\}\}},\,\,\,{\{\{\{2,-2\},\{-1,2\}\}\}},
{\{\{0,1,0\},\{1,0,1\}\}},\,{\{\{\{\{1,0\}\},1\}\}},\,{\{\{1,1\},\{1,0\}\}},\,{\{\{\{1,1\},1\},\{\{2,2\},1\}\}}}
\end{codeSyntax}
Note that the different representations are given in terms of their Dynkin coefficients (with the exception of $SU(2)$'s of the custodial group, which we label using the dimension) together with their multiplicity. Thus, \texttt{{\{\{\{1,0\}\},1\}}} refers to the $\bm{5}$ of $SO(5)$, with multiplicity 1; whereas \texttt{{\{\{1,1\},1\},\{\{2,2\},1\}\}}} indicates the $(1,1)+(\bm{2},\bm{2})$ of $SU(2)\times SU(2)$.

The first  projection matrix, in this case $P=\left(\begin{smallmatrix}0&1&0\\1&0&1\end{smallmatrix}\right)$, characterises how the subgroup $\mathcal{H}=SO(5)$ is embedded in $\mathcal{G}=SU(4)$. Specifically, $\omega' = P\omega$ for any pair of weights $\omega$ and $\omega'$ of $\mathcal{G}$ and $\mathcal{H}$, respectively~\cite{Fonseca:2020vke}. An analogous comment applies to the second projection matrix. 

In those cases where the subgroup $\mathcal{H}$ may contain one or more $U(1)$'s, namely when $\mathcal{H}=\mathcal{H}_{\textrm{non-ab}}\times U(1)^{n}$, rather than provide all trivial possibilities $n=0,1,\cdots$ as different entries in the list of models, we only print the one associated to the largest $n$.
For example, in the case of $SU(4)/SU(2)^2$ we may add at most one $U(1)$ to $\mathcal{H}$.
In the corresponding entry in the ancillary file,
the $U(1)$ is explicit in the name of $\mathcal{H}$, \texttt{{\{"SU2","SU2","U1"\}}}, as well as in the Cartan matrix, \texttt{{\{\{\{2\}\},\{\{2\}\},\{\}\}}}; and the number of singlets in the decomposition of $\bm{r}_\mathcal{H}$ under $SU(2)\times SU(2)$ is specified with a list \texttt{\{0,1\}} in \texttt{{\{\{\{1,1\},\{0,1\}\},\{\{2,2\},2\}\}}}. Hence these cosets yield either a two-Higgs-doublet model  (2HDM)~\cite{Mrazek:2011iu} or a 2HDM extended with a singlet~\cite{Sanz:2015sua,Chala:2018qdf}. 

One last comment, involving a subtlety, is in order. In our scan, we sometimes find different models associated to the same $\mathcal{G}$, $\mathcal{H}$, $\bm{r}_\mathcal{H}$ and $\bm{r}_{SU(2)\times SU(2)}$. Looking at the projections matrices, one sees that they are different but it is important to note that more than one such matrix can yield the same branching rules; as such, a visual comparison of projection matrices is not an adequate method of distinguishing embeddings. The explanation for what is going on is the following: (1) as stated previously, we use the criteria of linear equivalence to access whether or not two embeddings are the same; (2) this relies on checking the branching rules for all representations of the group of interest, but in practice it is enough to look at the fundamental representation; (3) in some cases, it can happen that two inequivalent embeddings lead to the same branching rules for the adjoint representation, which is essentially the information we provide in $\bm{r}_\mathcal{H}$ and $\bm{r}_{SU(2)\times SU(2)}$. However, in general other representations decompose differently so the apparently repeated models are in reality inequivalent. 
As an example, there is only one embedding of $\mathcal{H}=G_{2}\times SO(5)$ in $\mathcal{G}=SO(7)\times SU(4)$, under which the coset space transforms
	as $\bm{r}_{\mathcal{H}}=\left(\bm{7},1\right)+\left(1,\bm{5}\right)$.
	There are then two inequivalent embeddings of $SU(2)\times SU(2)$ in $\mathcal{H}$ such that $\bm{r}_{\mathcal{H}}\rightarrow\bm{r}_{SU(2)\times SU(2)}=2\times\left(\bm{3},1\right)+\left(\bm{2},\bm{2}\right)+2\times\left(1,1\right)$:
	in one of them we have $\left(\bm{7},1\right)\rightarrow2\times \left(\bm{3},1\right)+\left(1,1\right)$
	and $\left(1,\bm{5}\right)\rightarrow\left(\bm{2},\bm{2}\right)+\left(1,1\right)$,
	while for the other embedding the branching rules are $\left(\bm{7},1\right)\rightarrow\left(\bm{3},1\right)+\left(\bm{2},\bm{2}\right)$
	and $\left(1,\bm{5}\right)\rightarrow\left(\bm{3},1\right)+2\times\left(1,1\right)$.

Nonetheless, since the light scalars depend only on what happens to the adjoint representation, in the ancillary file we provide only one model for each of these collections of apparently equivalent cosets; we are happy to provide the rest upon request.\footnote{Without this reduction we obtain a total of 752 models, 110 more than the 642 we mention in the text and which we provide in the auxiliary file.}

\thinmuskip=2 mu
\medmuskip=3 mu
\thickmuskip=3 mu
\section{Highlighted results}
\label{sec:results}
\begin{table}[t!]
\begin{center}
\resizebox{\textwidth}{!}{
\begin{tabular}{cccc}
 \toprule
 $SU(2)\times SU(2)$ content & $\bm{r}_\mathcal{H}$ & $\mathcal{H}$ & $\mathcal{G}/\mathcal{H}$\\
 \hline\\[-0.3cm]
 \multirow{1}{*}{$(\bf{2},\bf{2})$} & $(\bm{2},\bm{2})$ & $SU(2)\times SU(2)$ & $SO(5)/SU(2)^2$~\cite{Agashe:2004rs}\\
 \hline\\[-0.3cm]
 \multirow{2}{*}{$(\mathbf{2},\mathbf{2})+(1,1)$} & $(\bm{2},\bm{2})+(1,1)$ & $SU(2)\times SU(2)$ & $SO(5)\times U(1)/SU(2)^2$~\cite{Gripaios:2016mmi}\\
 & $\bm{5}$ & $SO(5)$ & $SU(4)/SO(5)$~\cite{Gripaios:2009pe}\\
 \hline\\[-0.3cm]
 \multirow{4}{*}{$(\mathbf{2},\mathbf{2})+(1,\mathbf{3})$} & $(\bm{2},\bm{2})+(1,\bm{3})$ & $SU(2)\times SU(2)$ & $SU(2)\times SO(5)/SU(2)^2$\\
 & $\bm{7}$ & $G_2$ & $SO(7)/G_2$~\cite{Chala:2012af}\\
 & $\bm{7}$ & $SO(7)$ & $SO(8)/SO(7)$ \\
 & $(1,\bm{2},\bm{2})+(\bm{3},1,1)$ & $SU(2)^3$ & $SU(2)\times SU(2)\times SO(5)/SU(2)^3$\\
  \hline\\[-0.3cm]
 \multirow{7}{*}{$2\times (\mathbf{2},\mathbf{2})$} & $2\times(\bm{2},\bm{2})$ & $SU(2)\times SU(2)$ & $SU(4)/(SU(2)^2\times U(1))$~\cite{Mrazek:2011iu}\\
 & $\bm{4}+\bm{\bar{4}}$ & $SU(4)$ & $SU(5)/SU(4)\times U(1)$~\cite{Bertuzzo:2012ya}\\
 & $\bm{8}$ & $SO(7)$ & -- \\
 & $(\bm{2},\bm{4})$ & $SU(2)\times SO(5)$ & $Sp(6)/(SU(2)\times SO(5))$~\cite{Mrazek:2011iu}\\
 & $(1,\bm{2},\bm{2})+(\bm{2},1,\bm{2})$ & $SU(2)^3$ & -- \\
 & $\bm{8}_v$ & $SO(8)$ & $SO(9)/SO(8)$~\cite{Bertuzzo:2012ya}\\
 & $(1,\bm{2},1,\bm{2})+(\bm{2},1,\bm{2},1)$ & $SU(2)^4$ & $SO(5)^2/SU(2)^4$ \\ \hline
 \multirow{4}{*}{$(\mathbf{2},\mathbf{2})+2\times(1,1)$} & \multirow{2}{*}{$(\bm{2},\bm{2})+2\times (1,1)$} & \multirow{2}{*}{$SU(2)\times SU(2)$} & $SO(5)\times U(1)^2/SU(2)^2$ \\
 & & & $SU(2)\times SO(5)/(SU(2)^2\times U(1))$ \\
 & $1+\bm{5}$ & $SO(5)$ & $SU(4)\times U(1)/SO(5)$ \\
 & $\bm{6}$ & $SU(4)$ &  $SO(7)/SU(4)$~\cite{Chala:2016ykx,Balkin:2017aep}\\ \hline
 %
 \multirow{10}{*}{$(\mathbf{2},\mathbf{2})+(1,1)+(1,\mathbf{3})$} & $1+\bm{7}$ & $G_2$ & $SO(7)\times U(1)/G_2$\\
 & $(1,1)+(\bm{2},\bm{2})+(1,\bm{3})$  & $SU(2)^{2}$  & $SU(2)\times SO(5)\times U(1)/SU(2)^{2}$\\
 & $\bm{8}$ & $SO(7)$ & --\\
 & $1+\bm{7}$ & $SO(7)$ & $SO(8)\times U(1)/SO(7)$\\
 & $(\bm{2},\bm{4})$ & $SU(2)\times SO(5)$ & $Sp(6)/(SU(2)\times SO(5))$\\
 & $(1,\bm{5})+(\bm{3},1)$ & $SU(2)\times SO(5)$ & $SU(2)^2\times SU(4)/(SU(2)\times SO(5))$\\
 & $(1,\bm{2},\bm{2})+(\bm{2},\bm{2},1)$ & $SU(2)^3$ & --\\
 & $(1,1,1)+(1,\bm{2},\bm{2})+(\bm{3},1,1)$ & $SU(2)^3$ & $SU(2)^2\times SO(5)\times U(1)/SU(2)^3$\\
 & $\bm{8}_v$ & $SO(8)$ & $SO(9)/SO(8)$\\
 & $(1,1,\bm{2},\bm{2})+(\bm{2},\bm{2},1,1)$ & $SU(2)^4$ & $SO(5)^2/SU(2)^4$\\ \hline
 \multirow{7}{*}{$(\mathbf{2},\mathbf{2})+3\times(1,1)$} & \multirow{2}{*}{$(\bm{2},\bm{2})+3\times(1,1)$} & \multirow{2}{*}{$SU(2)^{2}$ } & $SO(5)\times U(1)^{3}/SU(2)^{2}$\\
 &  &  & $SU(2)\times SO(5)/SU(2)^{2}$\\
 
 & \multirow{2}{*}{$2\times 1 +\bm{5}$} & \multirow{2}{*}{$SO(5)$} & $SU(4)\times U(1)^2/SO(5)$ \\
 & & & $SU(2)\times SU(4)/(SO(5)\times U(1))$ \\
 & $1+\bm{6}$ & $SU(4)$ & $SO(7)\times U(1)/SU(4)$\\
 & $\bm{7}$ & $SO(7)$ & $SO(8)/SO(7)$\\
 & $(1,\bm{2},\bm{2})+(\bm{3},1,1)$ & $SU(2)^3$ & $SU(2)^2\times SO(5)/SU(2)^3$\\ \hline
 %
 \multirow{11}{*}{$(\mathbf{2},\mathbf{2})+4\times(1,1)$} & \multirow{3}{*}{$(\bm{2},\bm{2})+4\times(1,1)$} & \multirow{3}{*}{$SU(2)^{2}$} & $SO(5)\times U(1)^4/SU(2)^2$\\
 &  &  & $SU(2)\times SO(5)\times U(1)/SU(2)^{2}$\\
 & & & $SU(2)^2\times SO(5)/(SU(2)^2\times U(1)^2)$ \\
 & \multirow{2}{*}{$3\times1+\bm{5}$ } & \multirow{2}{*}{$SO(5)$ } & $SU(4)\times U(1)^3/SO(5)$ \\
 &  &  &  $SU(2)\times SU(4)/SO(5)$\\
 & \multirow{2}{*}{$2\times 1+\bm{6}$} & \multirow{2}{*}{$SU(4)$} & $SO(7)\times U(1)^2/SU(4)$\\
 & & & $SU(2)\times SO(7)/(SU(4)\times U(1))$\\
 & $1+\bm{7}$ & $SO(7)$ & $SO(8)\times U(1)/SO(7)$\\
 & $(1,\bm{5})+(\bm{3},1)$ & $SU(2)\times SO(5)$ & $SU(2)^2\times SU(4)/(SU(2)\times SO(5))$\\
 & $(1,1)+(1,\bm{5})+(\bm{2},1)$ & $SU(2)\times SO(5)$ & --\\
 & $(1,1,1)+(1,\bm{2},\bm{2})+(\bm{3},1,1)$ & $SU(2)^3$ & $SU(2)^2\times SO(5)\times U(1)/SU(2)^3$\\
& $2\times (\bm{2},1,1)+(1,\bm{2},\bm{2})$ & $SU(2)^3$ & $SO(5)\times SU(3)/(SU(2)^3\times U(1))$ \\
 & $\bm{8}_v$ & $SO(8)$ & $SO(9)/SO(8)$\\
 & $(1,1,\bm{2},\bm{2})+(\bm{2},\bm{2},1,1)$ & $SU(2)^4$ & $SO(5)^2/SU(2)^4$ \\
 %
 \bottomrule
\end{tabular}
}
\end{center}
\caption{\it CHMs with up to eight NGBs arranged according to the scalar field content. The symmetries on the scalar sector allowed from an IR perspective are specified. When relevant, citations to works in which these models have been studied are provided. For the five cases with a dash in the last column no CHM can be built; we include them in this table to support our discussion on this topic (see main text).}\label{tab:CHMswithN8}
\end{table}
Whilst the full list of all the CHMs with up to 13 NGBs is huge and hence only given in electronic form, the list of those models with up to eight NGBs is small enough to be provided in text. There are 44 in total, which we identify in Tab.~\ref{tab:CHMswithN8} (factoring out cases which differ only by $U(1)$'s, the number of models shrinks to 26). We group these CHMs in terms of their field content under the (custodial) SM.
Two immediate observations are in order. First, we note that:
\begin{codeSyntax}
For any choice of Standard Model field content, $\bm{r}_{SU(2)\times SU(2)}$,  there exists at least one composite Higgs model with the Nambu-Goldstone bosons transforming in that way.
\end{codeSyntax}
For example, the SM-group combination of a Higgs doublet and a real triplet can be realised via $SU(2)\times SO(5)/SU(2)^2$ as well as via $SO(7)/G_2$ or $SO(8)/SO(7)$ or even $SU(2)^2\times SO(5)/SU(2)^3$.\footnote{Note that $SU(2)$ is an active group in these examples; it is not spectating.} Although it might be surprising, this results extends not only to $m=13$ but actually to an arbitrary number of NGBs. In fact, any custodial-group field content can be realised in an $SO(m+1)/SO(m)$ CHM.

The validity of this last statement is rather simple to prove. An $m$-dimensional (potentially reducible) real, faithful and unitary representation of any group consists of a set of $m\times m$ orthogonal matrices, so they form a subgroup of the full set $O(m)$ of all such matrices. If the matrices have unit determinant, then they form a subgroup of $SO(m)$. Hence, for any $m$-dimensional real representation $\bm{r}_{SU(2)\times SU(2)}$ of $SU(2)\times SU(2)$ there is always  an embedding in $SO(m)$ such that the fundamental representation of this group transforms as $\bm{r}_{SU(2)\times SU(2)}$ under $SU(2)\times SU(2)$. Given that the coset space $SO(m+1)/SO(m)$ transforms as the fundamental representation of $SO(m)$, it can realise any (real) combination of $SU(2)\times SU(2)$ fields provided that this subgroup is appropriately embedded in $SO(m)$. 

In this way we obtain, for example, the CHM version of the Georgi-Machacek model~\cite{Georgi:1985nv}, composed of the light scalars $(\bm{2},\bm{2})+(\bm{3},\bm{3})$, which, to the best of our knowledge, has not been discussed previously in the literature. The corresponding coset is simply $SO(14)/SO(13)$, with the embedding of $SU(2)\times SU(2)$ into $SO(13)$ associated with the projection matrix $P=\left(\begin{smallmatrix}1&2&4&6&8&4\\1&0&2&2&0&1\end{smallmatrix}\right)$. It can be also realised with $SO(5)\times SO(10)/ (SU(2)^2\times SO(9))$, as well as with a more minimal coset $(SO(5)\times SU(4))/SU(2)^4$. Other scenarios not yet studied within the CHM context include the scalar minimal dark matter multiplets~\cite{Cirelli:2005uq}. These, together with all other single-field extensions of the Higgs doublet arising in composite models with up to $13$ NGBs are shown in Tab.~\ref{tab:singlefield}.
 \begin{table}[t!]
\begin{center}
\resizebox{0.95\textwidth}{!}{
\begin{tabular}{cccccc}
\hline\\[-0.3cm]
$\mathcal{G}$ & $\mathcal{H}$ & $\bm{r}_{\mathcal{H}}$ & $\bm{r}_{SU(2)\times SU(2)}$ \\
\hline\\
$SO(5)\times U(1)$ & $SU(2)\times SU(2)$ & $(1,1)+(\bm{2},\bm{2})$ & $(1,1)+(\bm{2},\bm{2})$ & \cite{Gripaios:2016mmi}\\
$SU(4)$ & $SO(5)$ & $\bm{5}$ & $(1,1)+(\bm{2},\bm{2})$ & \cite{Gripaios:2009pe}\\
$SU(4)$ & $SU(2)^2\times U(1)$ & $(\bm{2},\bm{2})_{-2}+(\bm{2},\bm{2})_{+2}$ & $2\times(\bm{2},\bm{2})$ & \cite{Mrazek:2011iu}\\
$SO(7)$ & $G_2$ & $\bm{7}$ & $(\bm{2},\bm{2})+(\bm{3},1)$ & \cite{Chala:2016ykx}\\
$SO(7)$ & $SU(2)^3$ & $(\bm{2},\bm{3},\bm{2})$ & $(\bm{2},\bm{2})+(\bm{4},\bm{2})$\\
$Sp(6)$ & $SU(2)\times SO(5)$ & $(\bm{4},\bm{2})$ & $2\times(\bm{2},\bm{2})$ & \cite{Mrazek:2011iu}\\
$SU(2)\times SO(5)$ & $SU(2)\times SU(2)$ & $(\bm{3},1)+(\bm{2},\bm{2})$ & $(\bm{2},\bm{2})+(\bm{3},1)$\\
$SU(5)$ & $SU(4)\times U(1)$ & $\bm{\overline{4}}_{-5}+\bm{4}_{+5}$ & $2\times(\bm{2},\bm{2})$ & \cite{Bertuzzo:2012ya}\\
$SO(8)$ & $SO(7)$ & $\bm{7}$ & $(\bm{2},\bm{2})+(\bm{3},1)$\\
$SO(9)$ & $SO(8)$ & $\bm{8_{v}}$ & $2\times(\bm{2},\bm{2})$ & \cite{Bertuzzo:2012ya}\\
$Sp(8)$ & $SU(2)\times Sp(6)$ & $(\bm{2},\bm{6})$ & $(\bm{2},\bm{2})+(\bm{4},\bm{2})$\\
$Sp(8)$ & $SU(2)\times Sp(6)$ & $(\bm{2},\bm{6})$ & $(\bm{2},\bm{2})+(\bm{4},\bm{2})$\\
$SO(5)\times SU(3)$ & $SU(2)^3$ & $(\bm{2},\bm{2},1)+(1,1,\bm{5})$ & $(\bm{2},\bm{2})+(\bm{5},1)$\\
$SO(5)^2$ & $SU(2)^3$ & $(\bm{7},1,1)+(1,\bm{2},\bm{2})$ & $(\bm{2},\bm{2})+(\bm{7},1)$\\
$SO(5)^2$ & $SU(2)^4$ & $(\bm{2},\bm{2},1,1)+(1,1,\bm{2},\bm{2})$ & $2\times(\bm{2},\bm{2})$\\
$G_2\times SO(5)$ & $SU(2)^4$ & $(\bm{4},\bm{2},1,1)+(1,1,\bm{2},\bm{2})$ & $(\bm{2},\bm{2})+(\bm{4},\bm{2})$\\
$SU(2)^2\times SO(5)$ & $SU(2)^3$ & $(\bm{3},1,1)+(1,\bm{2},\bm{2})$ & $(\bm{2},\bm{2})+(\bm{3},1)$\\
$SO(10)$ & $SO(9)$ & $\bm{9}$ & $(\bm{2},\bm{2})+(\bm{5},1)$\\
$SO(5)\times SU(4)$ & $SU(2)^2\times SO(5)$ & $(\bm{2},\bm{2},1)+(1,1,\bm{5})$ & $(\bm{2},\bm{2})+(\bm{5},1)$\\
$SO(5)\times SU(4)$ & $SU(2)^4$ & $(\bm{2},\bm{2},1,1)+(1,1,\bm{3},\bm{3})$ & $(\bm{2},\bm{2})+(\bm{3},\bm{3})$\\
$SO(5)\times SO(7)$ & $SU(2)^2\times G_2$ & $(\bm{2},\bm{2},1)+(1,1,\bm{7})$ & $(\bm{2},\bm{2})+(\bm{7},1)$\\
$SO(5)\times Sp(6)$ & $SU(2)^3\times SO(5)$ & $(\bm{2},\bm{2},1,1)+(1,1,\bm{2},\bm{4})$ & $(\bm{2},\bm{2})+(\bm{4},\bm{2})$\\
$SO(12)$ & $SO(11)$ & $\bm{11}$ & $(\bm{2},\bm{2})+(\bm{7},1)$\\
$SO(13)$ & $SO(12)$ & $\bm{12}$ & $(\bm{2},\bm{2})+(\bm{4},\bm{2})$\\
$SO(5)\times SO(8)$ & $SU(2)^2\times SO(7)$ & $(\bm{2},\bm{2},1)+(1,1,\bm{7})$ & $(\bm{2},\bm{2})+(\bm{7},1)$\\
$SO(5)\times SO(9)$ & $SU(2)^2\times SO(8)$ & $(\bm{2},\bm{2},1)+(1,1,\bm{8_{v}})$ & $(\bm{2},\bm{2})+(\bm{4},\bm{2})$\\
$SO(14)$ & $SO(13)$ & $\bm{13}$ & $(\bm{2},\bm{2})+(\bm{9},1)$\\
$SO(14)$ & $SO(13)$ & $\bm{13}$ & $(\bm{2},\bm{2})+(\bm{3},\bm{3})$\\
$SO(5)\times SO(10)$ & $SU(2)^2\times SO(9)$ & $(\bm{2},\bm{2},1)+(1,1,\bm{9})$ & $(\bm{2},\bm{2})+(\bm{9},1)$\\
$SO(5)\times SO(10)$ & $SU(2)^2\times SO(9)$ & $(\bm{2},\bm{2},1)+(1,1,\bm{9})$ & $(\bm{2},\bm{2})+(\bm{3},\bm{3})$\\
\hline
\end{tabular}
}
\end{center}
\caption{\it CHMs with at most 13 NGBs and which transform as a bi-doublet plus a single extra irrep of the custodial group.}\label{tab:singlefield}
\end{table}
From the tables, it is also clear that all composite singlet extensions of the scalar sector have been studied. For 2HDMs, we find one option that has not been considered previously in the literature, namely $SO(5)^2/SU(2)^4$, but this is just two minimal CHMs charged under different $SO(5)$'s and $SU(2)^2$'s with the custodial group being the diagonal subgroup of the two $SU(2)^2$'s. Extensions with one real triplet are much less explored; we find four possible cosets, one of which --- $SO(7)/G_2$ --- has been previously studied in Refs.~\cite{Chala:2012af,Ballesteros:2017xeg}.

A second observation is:
\begin{codeSyntax}
 Certain symmetries of the Nambu-Goldstone boson sector, acceptable from an IR perspective, are never realised within a composite Higgs model. Specifically, there are choices of $\mathcal{H}$ and  $\bm{r}_\mathcal{H}$ for which there is no larger group  $\mathcal{G}\supset\mathcal{H}$ such that  $\text{Adj}(\mathcal{G})\to\text{Adj}(\mathcal{H})+ \bm{r}_\mathcal{H}$.
\end{codeSyntax}
The clearest example is a 2HDM sector in which the eight scalar degrees of freedom transform as the $\bm{8}$ of $\mathcal{H}=SO(7)$. In CHMs, either the scalar sector is not that symmetric, or the symmetry is larger than that; $\mathcal{H}=SO(8)$ is a possibility, since  the coset space $SO(9)/SO(8)$ transforms as an octet of this group. Needless to say, the $\bm{8}$ of $SO(7)$ does not fulfil the closure condition.

We provide further information about the landscape of CHMs in Fig.~\ref{fig:plots}.
\begin{figure}[t]
 \includegraphics[width=0.5\columnwidth]{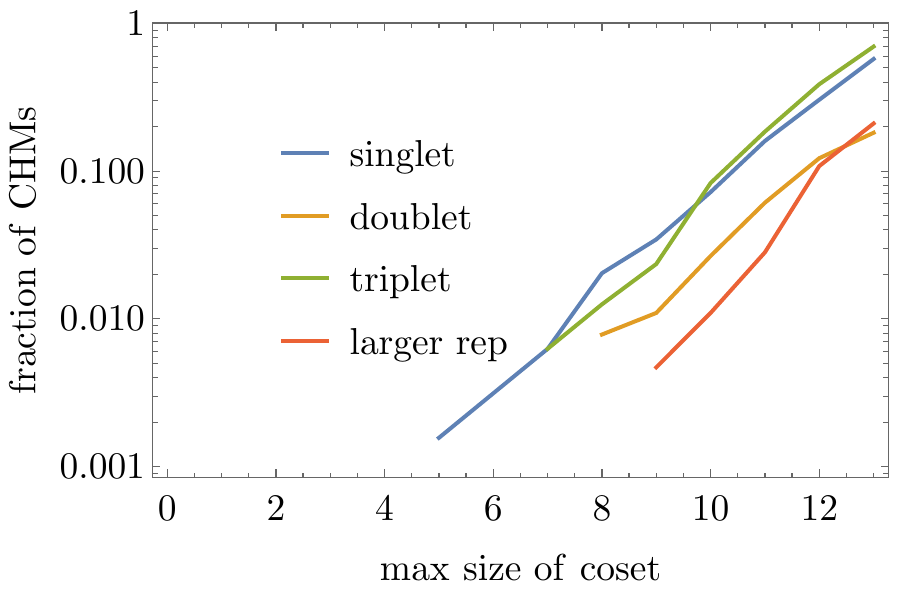}
 \includegraphics[width=0.5\columnwidth]{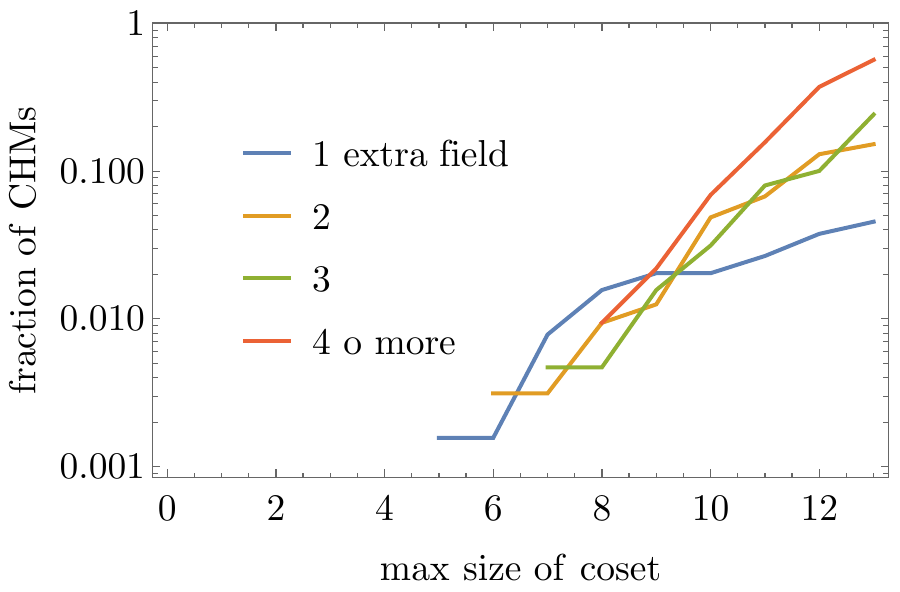}
 \includegraphics[width=0.5\columnwidth]{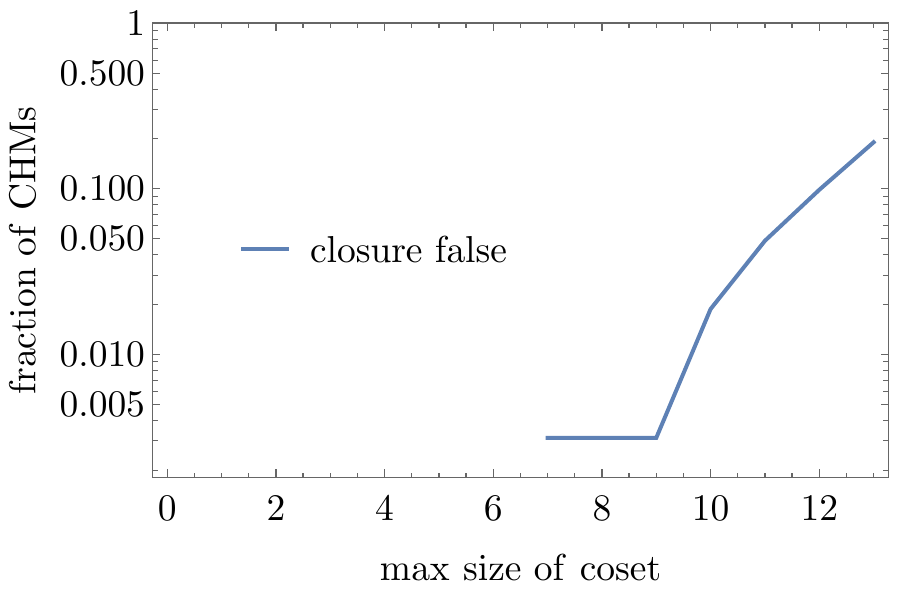}
 \includegraphics[width=0.5\columnwidth]{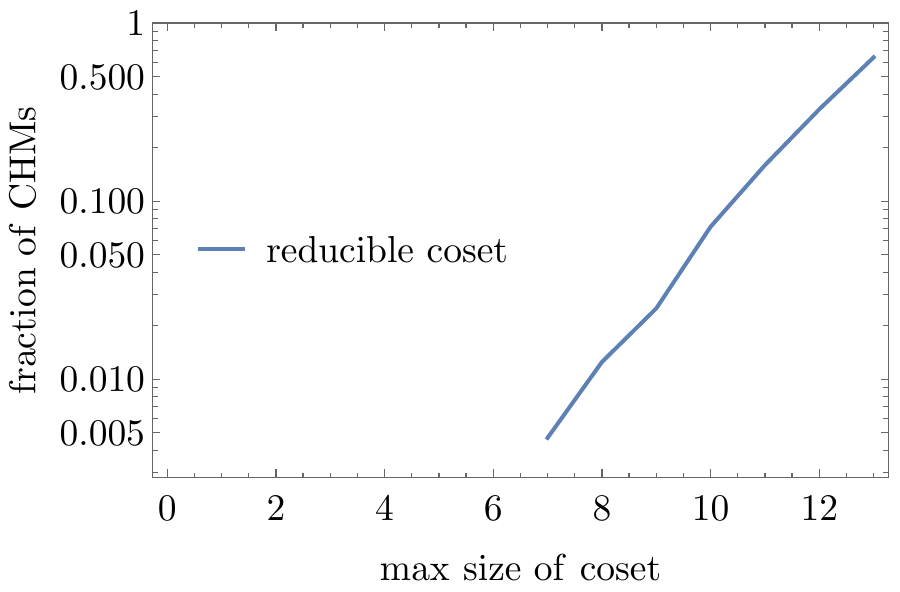}
 \caption{\it Top left) Fraction of CHMs with extra singlets, doublets, triplets and larger multiplets. Top right) Fraction of CHMs with different number of extra scalar multiplets. Bottom left) Fraction of CHMs which do not fulfil the closure condition. Bottom right) Fraction of CHMs in which $\bm{r}_{\mathcal{H}}$ is reducible. In all cases, we represent this information as a function of the maximum size of the coset space. }\label{fig:plots}
\end{figure}
In the top left panel we show the fraction of models involving different multiplets of the custodial group. Singlets and doublets are understood as $(1,1)$ and $(\bm{2},\bm{2})$ of $SU(2)\times SU(2)$, respectively; whereas triplets can be either $(1,\bm{3})$ or $(\bm{3},\bm{3})$ of $SU(2)\times SU(2)$. Larger multiplets comprise any other representations with one $SU(2)$ component of dimension 4 or higher. We note that, perhaps surprisingly, triplets are more common than doublets and singlets (about 70\% of the CHMs involve triplets), and in fact doublets are not as common as larger multiplets for big enough cosets.\footnote{We consider no $U(1)$ factors in $\mathcal{G}$ and a maximum number of them in $\mathcal{H}$. Singlets are more abundant otherwise.}
In the right panel of the figure, we show the fraction of cosets with different number of scalar irreps of the custodial group (besides the Higgs bi-doublet). With the exception of very small ones, clearly most cosets deliver more than four $SU(2)\times SU(2)$ multiplets. 

In the bottom left panel, we plot the fraction of CHMs which do not fulfill the closure condition. For simplicity of the calculation, we do not take the full landscape of 642 CHMs, but rather the sample of those with no abelian factors and in which $\bm{r}_{\mathcal{H}}$ does not involve complex components --- thus excluding, for example, $\bm{4}_{-5}+\bm{\overline{4}}_{+5}$ in $SU(5)/(SU(4)\times U(1))$. This reduces the sample of CHMs to 478. From the figure we see that for many, about 20\% of them, the closure condition fails, indicating that a big part of the CHMs are non-symmetric. (The only case for each both  $\mathcal{G}$ and  $\mathcal{H}$ are simple groups is $SO(7)/G_2$.) A different aspect of this same (smaller) sample of CHMs that we study is the fraction of those in which $\bm{r}_{\mathcal{H}}$ is reducible. This is important because it determines the number of free parameters in the non-linear sigma model; see the discussion below Eq.~\eqref{eq:lag}. We see that CHMs with one single parameter (such as $SO(m+1)/SO(m)$), which are the most studied in the literature, are actually rare.

\section{Conclusions}
\label{sec:conclusions}
We have made an exhaustive compilation of CHMs with at most 13 NGBs. In particular, we have computed all cosets $\mathcal{G}/\mathcal{H}$ with at most 13 generators, with $\mathcal{G}$ and $\mathcal{H}$ being compact semi-simple Lie groups. Factoring out $U(1)$'s whose only effect is to change the number of singlet NGBs, we have found 642 models. Along the way, a number of interesting results were derived. 

First, we have obtained conditions for a set of scalar fields to be UV completable into a (symmetric) CHM, although about 20\% of the CHMs in our scan do not fulfil this condition, implying that a large fraction of them are non-symmetric. As far as we can tell, and in line with previous works~\cite{Low:2014nga,Low:2014oga}, these models can not be reconstructed on the basis of IR data only. Instead, the full knowledge of both $\mathcal{H}$ and $\mathcal{G}$ is needed, which we have provided here.

Second, we have found that the set of CHMs is actually finite for any fixed number $m$ of NGBs. 
That is because the gap between the number of generators of a group $\mathcal{G}$ and those of its maximal subgroups grows as we consider larger and larger $\mathcal{G}$'s. For the particular case $m\leq13$, we find that $\mathcal{G}$ can have a rank of at most 8. For rank 3 we reproduce all the results of Ref.~\cite{Bellazzini:2014yua}, and extend these with the cosets $Sp(6)/(SU(2)\times SO(5))$, $SO(7)/SU(2)^3$ and $Sp(6)/SU(2)^3$, which provide one singlet, one bi-doublet and one triplet; one bi-doublet and one quartet; and one singlet, two bi-doublets and one triplet, respectively. The first two are not listed in Ref.~\cite{Bellazzini:2014yua} because they involve less Higgs doublets than in different embeddings of the custodial group into $\mathcal{H}$. The last one, in which $\bm{r}_{\mathcal{H}}$ is reducible, can be obtained from $Sp(6)$ via an intermediate breaking to $SU(2)\times SO(5)$.\footnote{Nevertheless we note here that in general the reducibility of $\bm{r}_{\mathcal{H}}$ is unrelated to the existence of an intermediate breaking (i.e., it is unrelated to whether or not $\mathcal{H}$ is a maximal subgroup of $\mathcal{G}$). For example, under the maximal subgroup $SU(2)$ of $Sp(6)$, the corresponding coset space transforms as $\mathbf{7}+\mathbf{11}$.} On top of these we find two models based on $\mathcal{G}=SU(2)\times SO(5)$, each of which can be seen as the combination of two separate cosets.

Third, we have found that any combination of multiplets of the custodial symmetry shows up as the NGB sector of some CHM. In general, there is more than one CHM with the same SM field content (but different dynamics). In this respect, it is worth highlighting that we have found three CHMs with real triplets whose phenomenology, to the best of our knowledge, has not been explored in the literature. These are $SO(8)/SO(7)$, $SU(2)^2\times SO(5)/SU(2)^3$ and $SU(2)\times SO(5)/SU(2)^2$. This observation, combined with our finding that triplet scalars are, together with singlets, the multiplets that arise most often in CHMs, further motivates the search for electroweak triplets at colliders and other facilities. Moreover we have found three different realizations of the Georgi-Machacek model~\cite{Georgi:1985nv} composed of the light scalars $(\bm{2},\bm{2})+(\bm{3},\bm{3})$, as well as realizations of the minimal dark matter candidates \cite{Cirelli:2005uq}.

Finally, we have also observed that some symmetries of the scalar sector, which are perfectly valid from the IR point of view, are never realised within CHMs. This occurs only for scalar sectors with at least four NGBs. For example, a 2HDM in which the eight real degrees of freedom transform in the $\mathbf{8}$ of $\mathcal{H}=SO(7)$ is not permitted within CHMs. This finding is related to the first point raised above, namely the fact that representations of this sort do not fulfil the closure condition.

Among other future directions, this work could be extended by considering more than 13 NGBs. A more enlightening extension would be to allow a non-custodial $\mathcal{H}$, requiring only that it contains the SM gauge group $SU(2)\times U(1)$, in which case large corrections to the $\rho$ parameter must be avoided by assuming that the compositeness scale $f$ is large. Likewise, we could consider those scenarios in which $(SU(2)\times U(1))^2\subset \mathcal{H}$, in application to little-Higgs models~\cite{Arkani-Hamed:2002ikv,Arkani-Hamed:2002iiv}.

\section*{Acknowledgments}
We thank Ian Low and Jose Santiago for useful discussions.
This work has received financial support from grants RYC2019-027155-I (Ramón y Cajal), PID2019-106087GB-C22, PID2021-128396NB-I00 and PID2022-139466NB-C22 funded by the MCIN /AEI/ 10.13039/501100011033, ``El FSE invierte en tu futuro'' and ``FEDER Una manera de hacer Europa'', as well as from  the Junta de Andalucía grants FQM 101 and P21-00199.

%

\bibliographystyle{style} 

\bibliography{refs} 

\end{document}